\documentclass[aip,pop,floats,reprint,superscriptaddress]{revtex4-1}
\usepackage{amssymb}
\usepackage{amsmath}
\usepackage[usenames]{color}
\usepackage{colortbl}
\usepackage{graphicx}

 \newcommand{\const}{\mbox{const}}
 \def\bc{\begin{center}}          \def\ec{\end{center}}
 \allowdisplaybreaks
\begin{document}
 \title{Radial equilibrium of relativistic particle bunches in plasma wakefield accelerators}
 \author{K.V.Lotov}
 \affiliation{Novosibirsk State University, 630090, Novosibirsk, Russia}
 \affiliation{Budker Institute of Nuclear Physics SB RAS, 630090, Novosibirsk, Russia}
 \date{\today}
 \begin{abstract}
Drive particle beams in linear or weakly nonlinear regimes of the plasma wakefield accelerator quickly reach a radial equilibrium with the wakefield, which is described in detail for the first time. The equilibrium beam state and self-consistent wakefields are obtained by combining analytical relationships, numerical integration, and first-principle simulations. In the equilibrium state, the beam density is strongly peaked near the axis, the beam radius is constant along the beam, and longitudinal variation of the focusing strength is balanced by varying beam emittance. The transverse momentum distribution of beam particles depends on the observation radius and is neither separable, nor Gaussian.
 \end{abstract}
 \maketitle

\section{Introduction}

Acceleration of particles in plasmas has been studied since 1980s as a possible way to future high energy colliders. There are two modifications of this concept, which differ in methods of driving the plasma wave: plasma wakefield acceleration (PWFA) driven by a charged particle beam\cite{PRL54-693} and laser wakefield acceleration (LWFA) driven by a short laser pulse.\cite{PRL43-267} In both cases, the accelerated particle bunch (witness) must travel for some distance in the plasma to gain a substantial energy. The witness resides on the slope of a potential well (accelerating bucket) that moves with the speed of light $c$ and, in addition to accelerating the bunch, also keeps it transversely focused.

The period of radial (betatron) oscillations of the bunch particles is universally much shorter than the acceleration distance.\cite{NIMA-410-461} The accelerated bunch thus has enough time to reach a radial equilibrium with the focusing field. The only exception is the strongly nonlinear regime (called blow-out or bubble regime\cite{PRA44-6189,APB74-355}), in which the focusing force is strictly linear in radius, and phase mixing of particle oscillations does not occur. Drive bunches in PWFA (that reside on decelerating slopes of potential wells) also reach a radial equilibrium shortly after injection into the plasma.

Finding an equilibrium bunch shape in a fixed potential well is a straightforward, though sometimes technically cumbersome, task. However, the potential well is often not fixed. The drive bunches in PWFA create the potential well themselves, so the shape of the well depends on the bunch shape, and vice versa. Similarly, the witness, which seemingly propagates in the fixed wakefield of the driver, can locally modify this wakefield if the witness charge is sufficiently high to provide wave loading.\cite{PRL88-014802,NatPhys5-363}

Self-consistent bunch equilibria are of importance for wakefield acceleration schemes in which particle bunches propagate in a linear or moderately nonlinear plasma wave. Interest to these schemes varied in the history. In early PWFA experiments\cite{PRL61-98,PRA39-1586,NIMA-292-12}, the density of available particle bunches was small, and finding beam equilibria was a hot problem\cite{PFB2-1376,PRE49-4407,PoP2-1326,AIP396-75,PoP5-785,NIMA-410-388}. Later, with the advent of blow-out\cite{PRA44-6189,PoP9-1845} and bubble\cite{APB74-355,Nat.431-535,Nat.431-538,Nat.431-541} regimes, weakly nonlinear waves were relegated to the background. Recently, self-modulating proton drivers\cite{PRL104-255003,PoP20-083119} and optimization of staged LWFA for collider applications\cite{PRST-AB13-101301,PRST-AB14-091301,AIP1299-3} have renewed interest in moderately nonlinear waves and beam equilibria.

Previous attempts to calculate the equilibrium bunch shape in the wakefield created by the bunch itself have not met with success.\cite{PFB2-1376,PoP2-1326,AIP396-75} The reason is that, as we show below, the equilibrium particle distribution in the phase space is neither separable, nor Gaussian, and, therefore, not tractable analytically. Here we combine appropriate simplifying assumptions, analytical relationships, numerical integration, and cross-checking by first-principle simulations to reach the goal.

In Sec.~\ref{s2}, we enumerate and justify the simplifying assumptions and write out the key formulae that we use to construct the equilibrium solution. In Sec.~\ref{s3}, we describe the test case that we simulate to benchmark our theory. Then in Sec.~\ref{s4} we find the equilibrium state of the beam, and in Sec.~\ref{s5} discuss special features of this state. Main findings are briefly summarized in Sec.~\ref{s6}.

\begin{table*}[t]
 \caption{ Driver emittances in PWFA experiments.}\label{t1}
 \begin{tabular}{lcccccccc}\hline
  Facility, reference & $n_0$, $\text{cm}^{-3}$ \quad & $c/\omega_p$, $\mu$m \quad & $\sigma_r$, $\mu$m \quad & $\gamma_b m_b/m$ \quad & $\alpha$ \quad & $\varepsilon_\text{eq}$, mm mrad \quad & $\varepsilon_0$, mm mrad \quad & $\varepsilon_\text{eq} / \varepsilon_0$ \\ \hline
  AATF (ANL, Argonne)\cite{PRA39-1586} & $7.3\times 10^{12}$ & 2000 & 1400 & 41 & 0.08 & 1300 & 300 & 4.5 \\
  KEK (Tsukuba)\cite{NIMA-292-12} & $4\times 10^{11}$ & 8400 & 1000 & 490 & 0.07 & 480 & 3 & 150 \\
  FACET (SLAC, Stanford)\cite{Nat.515-92} & $5\times 10^{16}$ & 24 & 30 & $4\times 10^{4}$ & 1 & 5000 & 360 & 15 \\
  FACET (SLAC, Stanford)\cite{Nat.524-442} & $8\times 10^{16}$ & 20 & 20 & $4\times 10^{4}$ & 1 & 3000 & 200 & 15 \\
  ATF (BNL, Brookhaven)\cite{PRL112-045001} & $5\times 10^{15}$ & 75 & 120 & 113 & 0.05 & 300 & 13 & 23 \\
  AWAKE (CERN, Geneva)\cite{NIMA-829-76} & $7\times 10^{14}$ & 200 & 200 & $7\times 10^{5}$ & 0.5 & 45 & 3.5 & 13 \\
  PITZ (DESY, Zeuthen)\cite{NIMA-740-74} & $1\times 10^{15}$ & 170 & 42 & 42 & 0.1 & 15 & 0.372 & 40 \\
  CLARA (Daresbury Lab.)\cite{NIMA-829-43} & $6.5\times 10^{15}$ & 65 & 20 & 490 & 0.25 & 50 & 1 & 50 \\
  \hline
 \end{tabular}
\end{table*}

\section{Assumptions and basic relationships}\label{s2}

We consider axisymmetric beams and use cylindrical coordinates $(r, \phi, z)$ and the co-moving coordinate $\xi=z-ct$, with $\vec{e}_z$ being the direction of beam propagation. The most important simplifying assumption is neglect of the initial transverse momentum of beam particles. To justify this, we compare initial beam emittances with rough estimates of expected equilibrium emittances in recently conducted or widely discussed PWFA experiments. We represent the focusing force $F_r$ acting on beam particles as a fraction of the force exerted by the ion background in the blowout regime:\cite{PRE69-046405}
\begin{equation}\label{e1}
    |F_r| \sim 2 \pi \alpha n_0 e^2 r,
\end{equation}
where the dimensionless parameter $\alpha \lesssim 1$ characterizes the focusing strength, $n_0$ is the unperturbed plasma density, and $e>0$ is the elementary charge. Beam particles of the mass $m_b$ and energy $W_b = \gamma_b m_b c^2$ make betatron oscillations with the frequency
\begin{equation}\label{e2}
    \omega_\beta \sim \sqrt{\frac{|F_r|}{\gamma_b m_b r}}.
\end{equation}
A beam of the root-mean-square radius $\sigma_r$, which stays in equilibrium with this focusing field, must have the normalized emittance
\begin{equation}\label{e3}
    \varepsilon_\text{eq} \sim \frac{\gamma_b \sigma_r^2 \omega_\beta}{c} \sim k_p \sigma_r^2 \sqrt{\frac{\gamma_b m \alpha}{2 m_b}},
\end{equation}
where we introduced the electron mass $m$ and the plasma wavenumber $k_p=\omega_p/c$ determined by the plasma frequency $\omega_p = \sqrt{4 \pi n_0 e^2/m}$. The estimated equilibrium emittances $\varepsilon_\text{eq}$ and initial beam emittances $\varepsilon_0$ are compared in Table~\ref{t1}. When calculating $\varepsilon_\text{eq}$, we make rough estimates for the parameter $\alpha$ on the basis of wave nonlinearity degree or from the maximum longitudinal wakefield, whatever is indicated in the reference. We see that reaching the radial equilibrium is typically accompanied by order(s) of magnitude emittance blow-up. Consequently, the initial emittance can be neglected when studying the established beam equilibrium.

We also assume that the plasma is linearly responding to the beam, which is true if the beam density $n_b$ is much lower than the plasma density: $n_b \ll n_0$. Only in this case, plasma wakefields and the beam density can be related analytically for arbitrary beam shapes.

For definiteness, we consider electron beams from now on. A highly relativistic electron propagating in the $z$-direction experiences the force $\vec{F}$, which is opposite to the gradient of the force potential $\Phi$:
\begin{equation}\label{e4}
    \vec{F} = -e (\vec{E} + [\vec{e}_z, \vec{B}]) = - \nabla \Phi.
\end{equation}
If the beam density is separable (as we assume),
\begin{equation}\label{e4a}
    n_b(r,\xi) = n_{b0} f(r) g(\xi),
\end{equation}
then the potential is also separable:\cite{PAcc20-171}
\begin{gather}
    \label{e6a} \Phi (r,\xi) = mc^2 \frac{n_{b0}}{n_0} R(r) G(\xi), \\
    \nonumber R(r) = -k_p^2 \int_0^r dr' r' I_1(k_p r') K_1(k_p r) f(r') \\
    \label{e7a}    - k_p^2 \int_r^\infty dr' r' I_1(k_p r) K_1(k_p r') f(r'),\\
    \label{e7b} G(\xi) = k_p \int_\xi^\infty d\xi' \sin \bigl( k_p (\xi'-\xi) \bigr) g(\xi'),
\end{gather}
where $I_1$ and $K_1$ are modified Bessel functions. Note that our definition of $\Phi$ is opposite in sign to the commonly used definition of the wakefield potential,\cite{RMP81-1229} and both $\Phi (r, \xi)$ and $R(r)$ are negative in potential wells. This makes discussions of potential wells more intuitive.

If the relativistic factor $\gamma_b$ of the electron, the shape of the potential well, and the longitudinal position of the electron with respect to the well change slowly as compared to the period of transverse electron oscillations, then the energy of transverse motion is conserved:
\begin{equation}\label{e5}
    W_\text{tr} = \frac{\gamma_b m v_r^2}{2} + \Phi(r,\xi) = \const,
\end{equation}
where $v_r$ is the radial velocity. If this electron makes transverse oscillations with the amplitude $r_a$, then its averaged contributions to the beam density at different radii do not depend on the depth of the potential well, but only on the well shape $R(r)$. Indeed, both the oscillation period
\begin{equation}\label{e6}
    \tau =  4 \int_0^{r_a} \frac{dr}{v_r} = 4 \int_0^{r_a} \frac{\sqrt{\gamma_b m} \, dr}{\sqrt{2 [\Phi(r_a)-\Phi(r)]}}
\end{equation}
and the time $dt$ that electron spends in a radial interval $dr$,
\begin{equation}\label{e7}
    dt = \frac{4 \sqrt{\gamma_b m} \, dr}{\sqrt{2 [\Phi(r_a)-\Phi(r)]}},
\end{equation}
identically scales with the well depth $\Phi(0)$. Consequently, we can assume that if the initial beam density is separable, then the equilibrium beam density is separable also, and the equilibrium radial profile of the beam is the same at all $\xi$. Early simulations of the long-term beam dynamics\cite{PoP5-785,NIMA-410-388,NIMA-410-461,EPAC98-806} agree with this assumption.

The oscillation amplitude distribution of beam particles $D(r_a)$ plays the key role in determination of the beam shape. We define it so that
\begin{equation}\label{e12a}
    \int_0^\infty D(r_a) \, dr_a = 1,
\end{equation}
and $D(r_a) dr_a$ is the fraction of beam particles that have amplitudes of transverse oscillations between $r_a$ and $r_a + dr_a$. Out of these particles, the fraction
\begin{equation}\label{e13}
    d^2 N_b = D(r_a) \, dr_a \, \frac{dt}{\tau} = \frac{D(r_a) \, dr_a \, dr}{\tilde\tau(r_a)\sqrt{R(r_a)-R(r)}}
\end{equation}
with
\begin{equation}\label{e13a}
    \tilde\tau(r_a) = \int_0^{r_a} \frac{dr}{\sqrt{R(r_a)-R(r)}}
\end{equation}
is currently within the radial interval between $r$ and $r+dr$. Integrating \eqref{e13} over all possible amplitudes yields $dN_b$, the fraction of particles in this radial interval, and the beam density profile:
\begin{equation}\label{e14}
    f(r) = \sigma_r^2 \frac{dN_b}{r \, dr} = \frac{\sigma_r^2}{r} \int_r^\infty \frac{D(r_a) \, dr_a}{\tilde\tau(r_a)\sqrt{R(r_a)-R(r)}}.
\end{equation}
This expression assumes $f(r)$ is normalized to give $2 \pi \sigma_r^{2}$ upon integrating across the beam.

Different amplitude distributions result in different beam shapes, so there is much freedom in how the equilibrium beam can look like. Here we aim at finding the most frequently encountered equilibrium, which low-emittance bunches evolve to.

Since the initial transverse velocity of beam particles is negligible, all particles are initially at their maximum radii, and
\begin{equation}\label{e12}
    D_0(r_a) = r_a f_0(r_a)/\sigma_r^2,
\end{equation}
where the subscripts `0' denote the initial beam state. As the beam starts to evolve, the above relation between the amplitude distribution $D(r_a)$ and the beam radial profile $f(r)$ breaks. As we show below, both the amplitude distribution and the radial profile change during transition to the equilibrium. However, the established amplitude distribution can be related to the initial one with a simple law obtained from simulations.

\section{Numerical benchmarking}\label{s3}

Our calculations are based on several assumptions, correctness of which needs to be checked by comparison of their consequences with numerical simulations. The simulated test case must contain all physics of beam transverse equilibration in its purest form and a minimum of other effects involved. The beam energy must be high to ensure a large difference between timescales of transverse dynamics and beam depletion, to comply with the requirement $\gamma_b = \const$. The initial angular spread of the beam must be low compared with the spread gained during equilibration. The beam density must be much lower than the plasma density to ensure linearity of the plasma response. The latter requirement conflicts with the low angular spread, as a cold beam, when focused by the plasma, creates a density singularity on the axis. Therefore, in simulations, the condition $n_b \ll n_0$ will break near the axis, but this will have no strong effect on the beam equilibrium, if the ratio $n_b/n_0$ is initially small and, consequently, the near-axis density spike is narrow.

\begin{table}[t]
 \caption{ Test case parameters.}\label{t2}
 \bc\begin{tabular}{ll}\hline
  Parameter, notation & Value \\ \hline
  Bunch density, $n_{b0}$ & $0.02 n_0$ \\
  Bunch length, $L$ & $2 \pi k_p^{-1}$ \\
  Bunch radius, $\sigma_{r}$ & $1.0\, k_p^{-1}$ \\
  Bunch energy, $W_b$ & $10^5 mc^2$ \\
  Bunch angular spread  & $10^{-6}$ \\
  Bunch energy spread  & 0 \\
  Simulated evolution time, $t_\text{max}$ & $10^5 \omega_p^{-1}$ \\
  Simulation time step & $10\, \omega_p^{-1}$ \\
  Simulation grid size (equal in $r$ and $\xi$) & $0.01 k_p^{-1}$ \\
  Number of simulated beam macro-particles \quad & $2 \pi \times 10^6$ \\
  \hline
 \end{tabular}\ec
\end{table}

In view of the foregoing, we take the test case parameters listed in Table~\ref{t2} and the initial beam shape
\begin{gather}
\label{e17}
    n_b(r,\xi) = n_{b0} f_0(r)
    \begin{cases}
        1,\quad & -L < \xi<0, \\
        0, & \text{otherwise},
    \end{cases} \\
\label{e17a}
    f_0(r) = \exp \left( -\frac{r^2}{2 \sigma_r^2} \right).
\end{gather}

Once the beam shape is defined, we can calculate the reference scale of transverse particle dynamics, as which we take the commonly used estimate $\omega_{\beta 0}$ of the betatron frequency. By analogy with \eqref{e2}, we substitute the initial Gaussian shape \eqref{e17}, \eqref{e17a} into the linear response formulae \eqref{e6a}--\eqref{e7b} and take the limit of small $r$:
\begin{equation}\label{e24}
    \omega_{\beta 0}^2 = \frac{1}{\gamma_b m} \lim_{r \to 0} \frac{1}{r} \frac{\partial \Phi}{\partial r}
    = \frac{\omega_p^2 n_{b0}}{2 \gamma_b n_0} A_0 G(\xi),
\end{equation}
where
\begin{equation}\label{e25}
    A_0 = 1 - \int_0^\infty K_0 (r) e^{-r^2/(2 k_p^2 \sigma_r^2)} r\, dr.
\end{equation}
For $k_p \sigma_r = 1$, the constant $A_0 \approx 0.54$. At the cross-section of strongest focusing, we have
\begin{equation}\label{e20a}
    k_p \xi = -\pi, \qquad G(\xi)=2, \qquad \omega_p/\omega_{\beta 0} \approx 3000,
\end{equation}
so the particles at this cross-section make about $t_\text{max} \omega_{\beta 0} /(2 \pi) \approx 5$ transverse oscillation during the simulation time.

\begin{figure}[tbh]
\includegraphics[width=231bp]{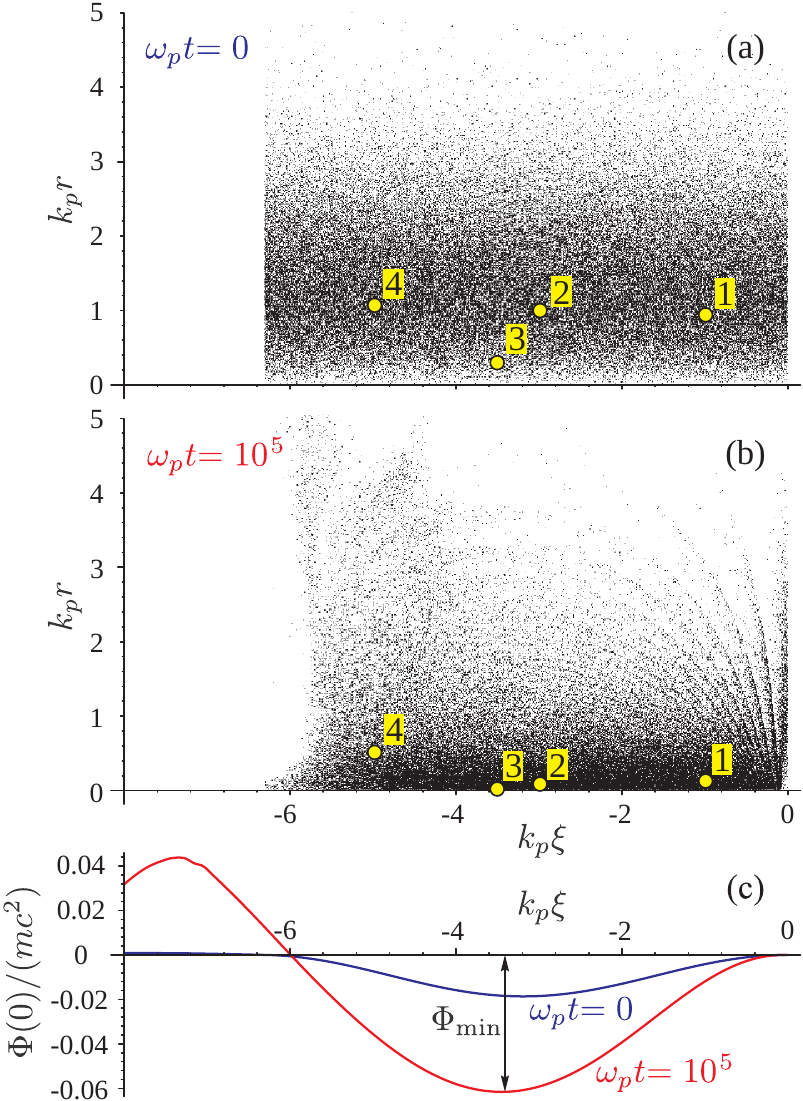}
\caption{ Beam portraits in real space at the beginning of interaction (a) and after the equilibrium is reached (b); the corresponding depths of the potential well (c). Particles followed in Fig.\,\ref{fig2-integrals} are marked by small circles and numbers in (a) and (b), and the maximum depth of the potential well $\Phi_\text{min}$ is indicated in (c) for the established equilibrium.}\label{fig1-beam}
\end{figure}

Simulations of the test case are made with two-dimensional fully kinetic quasistatic code LCODE\cite{PRST-AB6-061301,NIMA-829-350}. The initial and established states of the beam are illustrated in Fig.\,\ref{fig1-beam}.

\section{Equilibrium state}\label{s4}

In the equilibrium, the beam density profile $f(r)$ and the potential well shape $R(r)$ must simultaneously satisfy equations \eqref{e7a} and \eqref{e14}. The amplitude distribution $D(r_a)$ enters Eq.~\eqref{e14} and thereby determines a particular type of the equilibrium. The initial Gaussian density distribution \eqref{e17a} is not an equilibrium one, as is straightforward to check.

The equilibrium state can be found iteratively. We start from some amplitude distribution $D(r_a)$ and the initial beam shape defined by Eq.~\eqref{e12}. Then we calculate $R(r)$ from \eqref{e7a}, substitute it to \eqref{e14}, find $f(r)$, and so on. The scheme quickly converges, and there are no visual differences between the third iteration and further ones.

\begin{figure}[tbh]
\includegraphics[width=213bp]{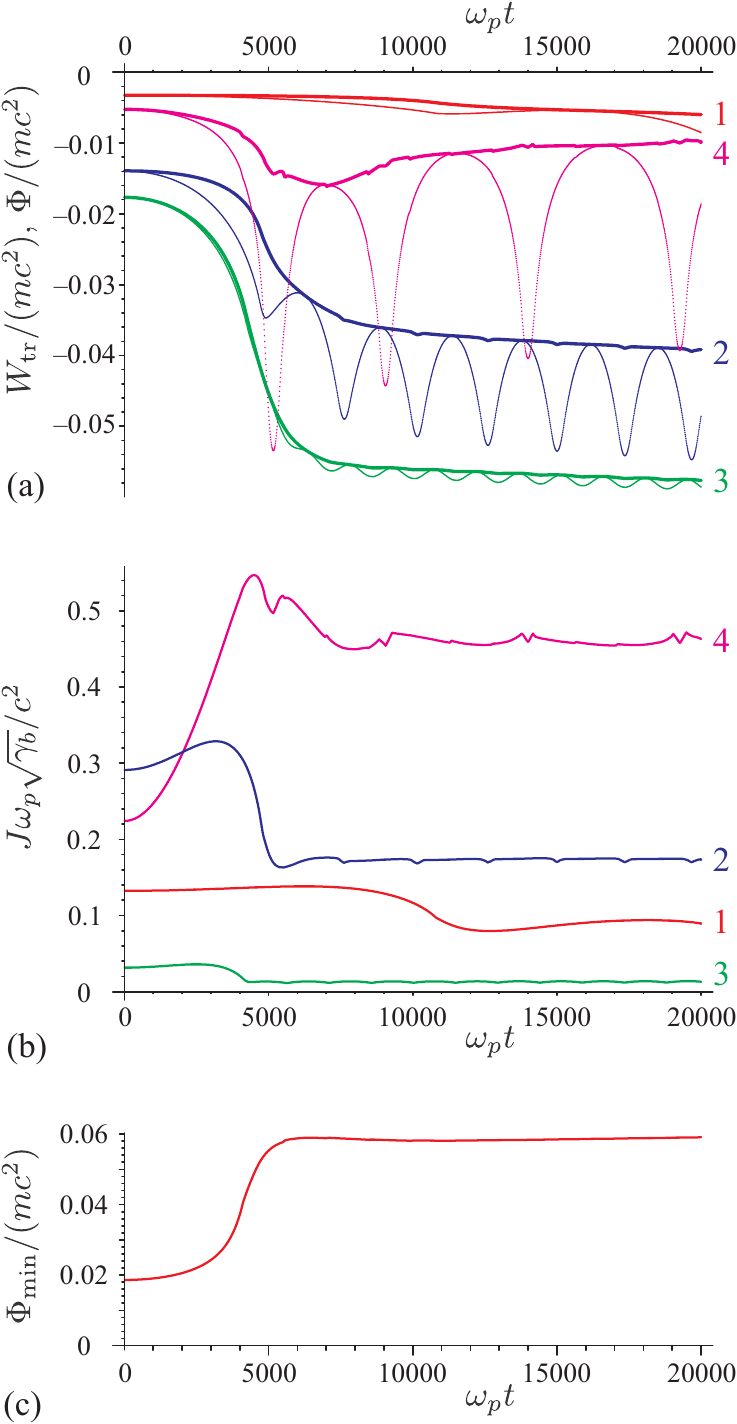}
\caption{ Simulated time dependencies of the transverse energy $W_\text{tr} (t)$ (thicker lines), the potential energy $\Phi(r(t))$ (thinner lines) (a) and the orbit phase area $J(t)$ (b) for several particles indicated in Fig.\,\ref{fig1-beam}(a) by corresponding numbers; the maximum depth of the potential well $\Phi_\text{min} (t)$ (c).}\label{fig2-integrals}
\end{figure}

However, the equilibrium state that follows from substituting the initial beam shape \eqref{e17a} into Eq.~\eqref{e12} does not agree with simulations of the test case. The reason is that the amplitude distribution changes during establishment of the equilibrium. Beam equilibration occurs at the time of several betatron oscillations, so the quantities, conservation of which usually helps to describe particle oscillations in slowly changing potential wells, are not constant. As illustrations, we show in Fig.\,\ref{fig2-integrals}(a) and (b) the transverse energy $W_\text{tr}$, the potential energy $\Phi$, and the area $J$ enclosed by the particle orbit in the phase space for several beam particles. Figure~\ref{fig2-integrals}(c) shows the time dependence of the well depth $\Phi_\text{min}$ at the very bottom, from which the timescale of equilibration is seen for comparison. All curves in Fig.\,\ref{fig2-integrals} are obtained from simulations. To find the orbit phase area at an arbitrary time $t$, we calculate the integral
\begin{equation}\label{e18}
    J = \oint v_r(r') dr' = 4 \int_0^{r_a} \sqrt{v_r^2(t) + \frac{2 [\Phi(r(t))-\Phi(r')]}{\gamma_b m} }\, dr'
\end{equation}
with $r_a$ defined by the equality
\begin{equation}\label{e19}
    v_r^2(t) + \frac{2 [\Phi(r(t))-\Phi(r_a)]}{\gamma_b m} = 0,
\end{equation}
as if the potential well would frieze at $t$, and the particle would make a full oscillation period in the frozen well starting from its position $r(t)$ with its radial velocity $v_r (t)$.

\begin{figure}[t]
\includegraphics[width=172bp]{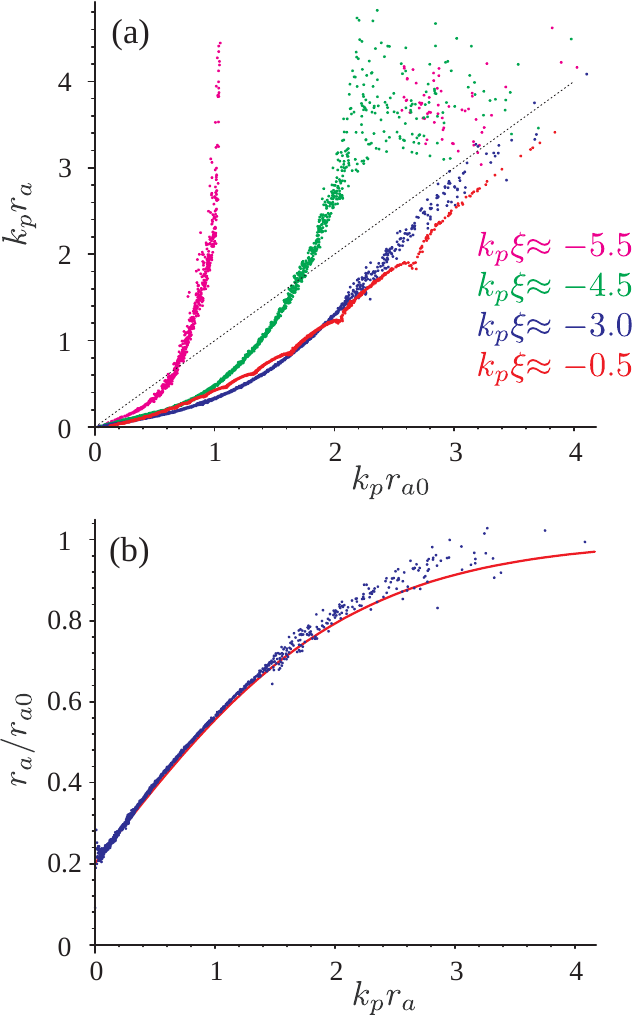}
\caption{ (a) Dependence of the equilibrium oscillation amplitude $r_a$ on the initial radius $r_{a0}$ for groups of particles located at several beam cross-sections (indicated at the figure by different colors). (b) The ratio $r_a/r_{a0}$ for particles with various oscillation amplitudes $r_a$ and $k_p \xi \approx -3$ (dots) and its approximation \eqref{e21} (line).}\label{fig3-amplitudes}
\end{figure}

Numerical simulations of the test case help to find the final amplitude distribution. It turns out that the relationship between the equilibrium amplitude $r_a$ and the initial position $r_{a0}$ of particles is the same in all cross-sections of the leading half of the beam, that is, in the interval $0 < k_p |\xi| \lessapprox \pi$ [Fig.\,\ref{fig3-amplitudes}(a)]. This relationship may be approximated as
\begin{equation}\label{e21}
    r_a/r_{a0} = B(r_a) \equiv 1 - \beta_1 \bigl(1 - \tanh(k_p r_a/\beta_2)\bigr)
\end{equation}
with $\beta_1 \approx 0.8$, $\beta_2 \approx 0.21$ [Fig.\,\ref{fig3-amplitudes}(b)]. The choice of these constants is determined by behavior of near-axis beam particles and does not depend on the initial beam radius $\sigma_r$. Consequently, these constants are the same for any beam. With the approximation \eqref{e21}, we can calculate the equilibrium amplitude distribution
\begin{equation}\label{e22}
    D(r_a) = D_0 (r_{a0} (r_a)) dr_{a0}/dr_a
\end{equation}
and then iteratively find the equilibrium beam density profile $f(r)$ and the potential well shape $R(r)$ in the leading half of the beam.

\begin{figure}[tbh]
\includegraphics[width=200bp]{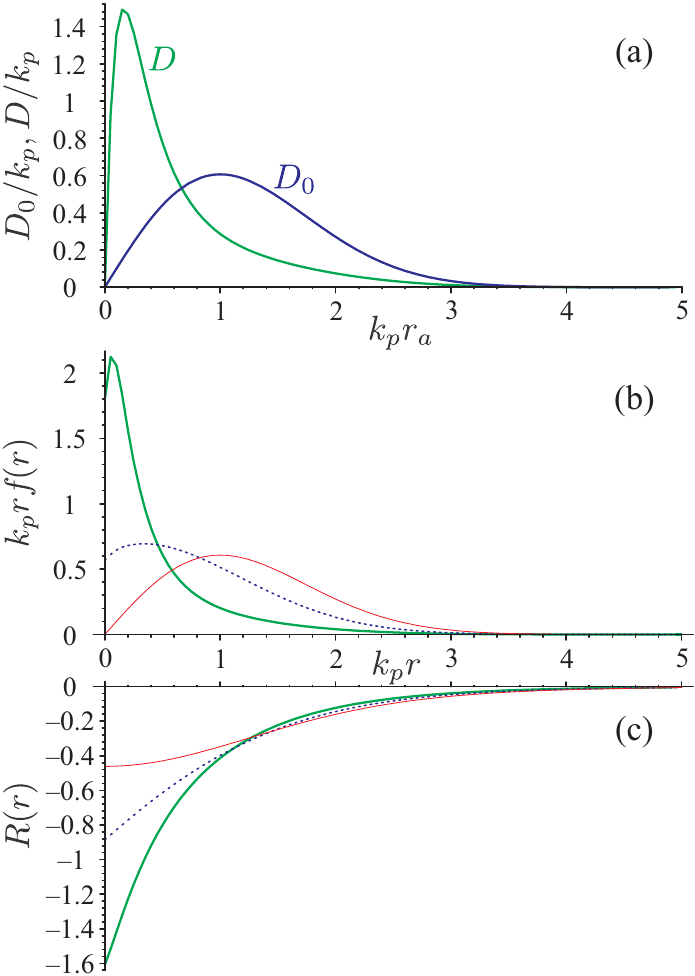}
\caption{ (a) The initial amplitude distribution $D_0$ and the amplitude distribution $D$ that follows from Eq.~\eqref{e21}. Radius-weighted density profiles (b) and shapes of the potential well (c) for the initial beam shape \eqref{e17a} (thin red lines), equilibrium beam with the initial amplitude distribution \eqref{e12} (dotted blue lines), and actual equilibrium beam (thick green lines). }\label{fig4-equilibrium}
\end{figure}

Initial and equilibrium amplitude distributions are compared in Fig.\,\ref{fig4-equilibrium}(a). Up to five-fold reduction of oscillation amplitudes for near-axis particles [Fig.\,\ref{fig3-amplitudes}(b)] results in strong peaking of the equilibrium amplitude distribution at small radii. For both distributions,
the equilibrium beam density has a $1/r$ singularity near the axis [Fig.\,\ref{fig4-equilibrium}(b)]. Equilibrium potential wells have no singularities and are funnel-shaped unlike the potential well of the Gaussian beam [Fig.\,\ref{fig4-equilibrium}(c)]. The constant derivative of the equilibrium potential near the axis means the radial electric field [which makes the dominant contribution to the radial force in Eq.~\eqref{e4}] is also constant there. The latter observation may be important for interpretation of numerical\cite{PRE68-047401,PoP10-2022} and real\cite{PRST-AB9-101301,Nat.445-741} PWFA experiments, in which the plasma is created through ionization of a neutral gas by the electric field of the beam.

The equilibrium state illustrated by Fig.\,\ref{fig4-equilibrium} is calculated for beams which initially have $k_p \sigma_r = 1$. For beams of different initial radii, the curves would have different radial scales and slightly different shapes, but their behavior at small $r$ ($k_p r \ll 1$) is qualitatively the same.

\section{Properties of the equilibrium state}\label{s5}

In this Section, we compare the calculated equilibrium state and its consequences with numerical simulations of the test case and commonly used estimates. We also discuss the origin of differences.

\begin{figure}[t]
\includegraphics[width=197bp]{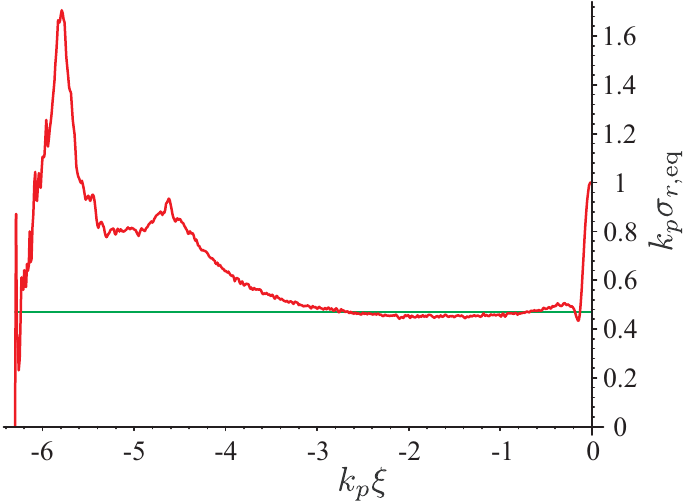}
\caption{ Simulated equilibrium radius of the test beam at $\omega_p t = 10^5$. The horizontal line shows the calculated value.}\label{fig5-radius}
\end{figure}

The root-mean-square (rms) radius of the equilibrium beam is
\begin{equation}\label{e22}
    \sigma_{r, \text{eq}} = \frac{1}{\sigma_r \sqrt{2}}\left( \int_0^\infty r^3 f(r) \, dr \right)^{1/2} \approx 0.47 \sigma_r,
\end{equation}
that is, twice smaller than the initial value. Simulations are consistent with this result, but only at the leading part of the beam (Fig.\,\ref{fig5-radius}). Also, the very head of the beam has a wider radius, as it has not reached the equilibrium during the simulated time period.

\begin{figure}[b]
\includegraphics[width=224bp]{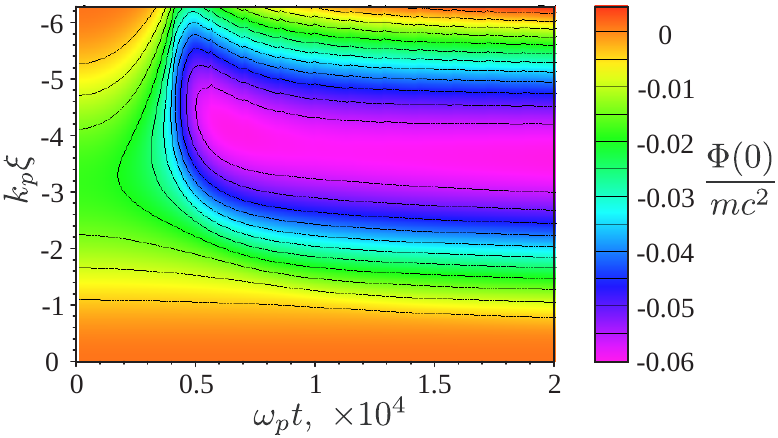}
\caption{ Time dependence of the on-axis potential $\Phi(0)$ at different beam cross-sections characterized by the co-moving coordinate $\xi$.}\label{fig6-potential}
\end{figure}

The qualitative difference of equilibrium states for leading and trailing parts of the simulated bunch comes from the non-monotonic behavior of the potential well depth at the trailing part (Fig.\,\ref{fig6-potential}). For $k_p |\xi| \gtrapprox \pi$, the potential well first deepens at approximately the same time it takes for beam particles to reach the axis [Fig.\,\ref{fig2-integrals}(a)] and then becomes shallower. When moving towards the axis for the first time, particles gain a substantial transverse momentum, which results in large-amplitude oscillations in the shallower well [Fig.\,\ref{fig3-amplitudes}(a)] and large equilibrium radius of the beam [Fig.\,\ref{fig1-beam}(b), Fig.\,\ref{fig5-radius}]. For $k_p |\xi| \lessapprox \pi$, the potential well deepens monotonically, and the force that accelerates particles towards the axis never exceeds the force that slows them down at the opposite side.

\begin{figure}[t]
\includegraphics[width=216bp]{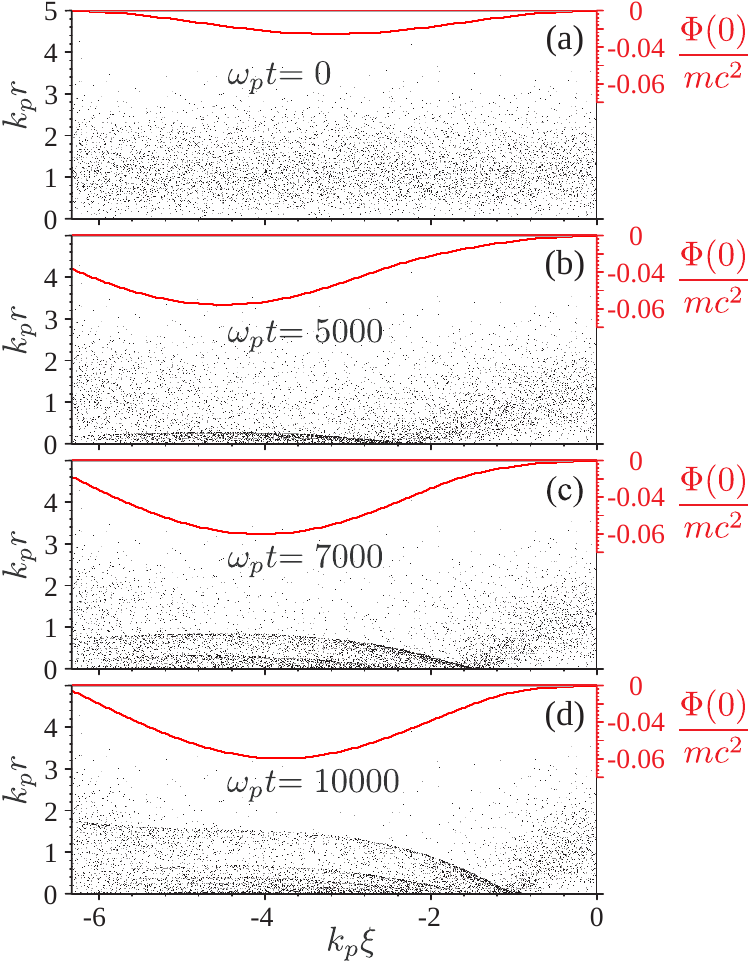}
\caption{ Simulated beam portraits (dots) and longitudinal profiles of potential wells (lines) at several times.}\label{fig7-compression}
\end{figure}

The non-monotonic behavior of the potential well depth follows from different beam pinching times at different cross-sections and from the quarter-period lag of the potential with respect to the beam [see Eq.~\eqref{e7b}]. The beam first pinches at the cross-section of strongest focusing, which is initially at $k_p |\xi| \approx \pi$ for our test bunch [Fig.\,\ref{fig7-compression}(a)]. This increases the focusing force at $k_p |\xi| \gtrapprox \pi$ [Fig.\,\ref{fig7-compression}(b)]. Later, the upstream beam parts located at $k_p |\xi| \lessapprox \pi$ pinch [Fig.\,\ref{fig7-compression}(c),(d)], and the deepest part of the potential well returns to its initial position thus causing shallowing of the well at $k_p |\xi| \gtrapprox \pi$. This backward-and-forward excursion of the potential minimum is possible only in regions of a negative derivative $\partial \Phi/ \partial \xi$, that is, in the accelerating phase of the wakefield. Therefore, the equilibrium found is typical for particle drivers, while for witnesses it may be different.

\begin{figure}[tbh]
\includegraphics[width=202bp]{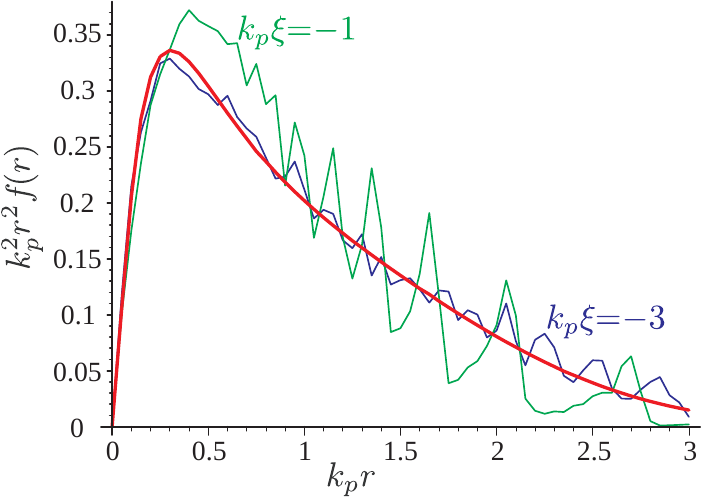}
\caption{ Calculated (thick red line) and simulated (thin lines) $r^2$-weighted transverse density profiles for two beam cross-sections at $\omega_p t = 10^5$.}\label{fig8-density}
\end{figure}

Because of the singular density behavior, it is more convenient to compare beam densities multiplied by $r^2$. Figure~\ref{fig8-density} shows the simulated beam density at two cross-sections and the calculated radial distribution of the beam density. The agreement is quite good, except for short scale rippling observed at the simulated profiles because of incomplete phase mixing of transverse oscillations.

\begin{figure}[b]
\includegraphics[width=195bp]{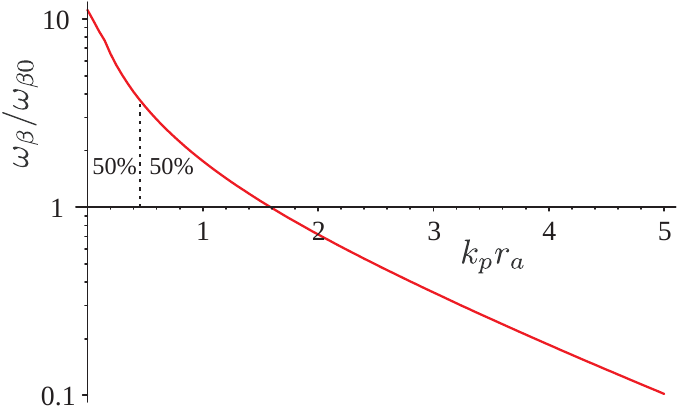}
\caption{ The frequency $\omega_\beta$ of transverse particle oscillations in the equilibrium as a function of the oscillation amplitude $r_a$. Half of the particles have oscillation amplitudes below the value indicated by the vertical dashed line.}\label{fig9-frequency}
\end{figure}

Figure~\ref{fig9-frequency} shows the frequency of particle betatron oscillations in equilibrium $\omega_\beta = 2 \pi/ \tau$ in comparison with the expression \eqref{e24}, where the period $\tau$ is calculated according to the formula \eqref{e6}. The difference in frequencies is so large that we use the logarithmic scale to show everything clearly. As most of the particles oscillate with amplitudes $r_a \ll k_p^{-1}$ [Fig.\,\ref{fig4-equilibrium}(a)], these particles also have oscillation frequencies several times higher than $\omega_{\beta 0}$.

For calculating the normalized emittance $\epsilon_\text{eq} (\xi)$ of the equilibrium beam, we use the formula
\begin{equation}\label{e28}
    \epsilon_\text{eq}^2 = \frac{\langle r^2 \rangle
\left( \langle p_r^2 \rangle + \langle p_\phi^2 \rangle \right) -
\langle r p_r \rangle^2}{2 m^2 c^2}.
\end{equation}
Here $p_r$ and $p_\phi$ are radial and azimuthal components of particle momentum, angle brackets denote averaging over the beam cross-section, and the factor of 2 in the denominator reduces the value to the commonly used single-coordinate ($x$ or $y$) emittance. In semi-analytical calculations, we put $\langle p_\phi^2 \rangle = 0$ and $\langle r p_r \rangle=0$. When analyzing simulations, all terms are taken into account.

\begin{figure}[tbh]
\includegraphics[width=195bp]{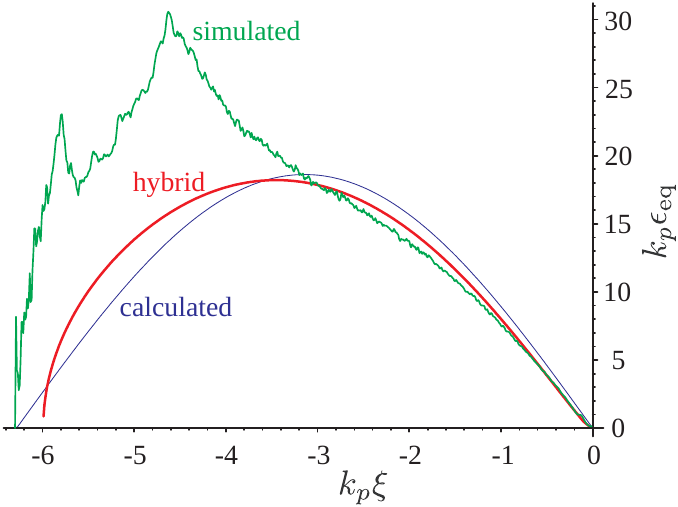}
\caption{ Equilibrium emittance of beam slices obtained from simulations (``simulated'') and from Eq.~\eqref{e31} with $G(\xi)$ defined either by Eq.\eqref{e7b} (``calculated''), or taken from simulations (``hybrid'').}\label{fig10-emittance}
\end{figure}

Derivation of the average radial momentum is the most complicated part of emittance calculation. For particles located at $r$ and having the oscillation amplitude $r_a$, we find from \eqref{e5}
\begin{equation}\label{e29}
    p_r^2 = 2 \gamma_b m [\Phi(r_a, \xi)-\Phi(r, \xi)].
\end{equation}
Then we multiply $p_r^2$ by the number of these particles \eqref{e13} and integrate over all locations and amplitudes:
\begin{multline}\label{e30}
    \langle p_r^2 \rangle = \int_0^\infty dr \int_r^\infty dr_a \frac{2 \gamma_b m [\Phi(r_a, \xi)-\Phi(r, \xi)] D(r_a)}{\tilde\tau(r_a)\sqrt{R(r_a)-R(r)}} \\
    = p_0^2 \int_0^\infty dr \int_r^\infty \frac{dr_a \sqrt{R(r_a)-R(r)} D(r_a)}{\tilde\tau(r_a)}
    = p_0^2 P_0,
\end{multline}
where
\begin{equation}\label{e31a}
    p_0 = mc \sqrt{2 \gamma_b n_{b0} G(\xi) / n_0}
\end{equation}
is the natural scale of transverse beam momentum at this cross-section, and the constant $P_0 \approx 0.2$ in our case. Finally,
\begin{equation}\label{e31}
    \epsilon_\text{eq} = \frac{\sqrt{\sigma_{r, \text{eq}}^2 \langle p_r^2 \rangle}}{mc} \approx 0.3 \sigma_r \sqrt{\frac{\gamma_b n_{b0} G(\xi)}{n_0}}.
\end{equation}
The estimated emittance obtained by substituting the betatron frequency \eqref{e24} into the expression \eqref{e3} is 75\% higher:
\begin{multline}\label{e32}
    \varepsilon_\text{eq} = \frac{\gamma_b \sigma_r^2 \omega_{\beta0}}{c} = k_p \sigma_r^2 \sqrt{\frac{A_0 \gamma_b n_{b0} G(\xi)}{2 n_0}} \\
    \approx 0.52 \sigma_r \sqrt{\frac{\gamma_b n_{b0} G(\xi)}{n_0}}.
\end{multline}
The emittance calculated from the formula \eqref{e31} is shown in Figure~\ref{fig10-emittance} in comparison with simulations. The longitudinal dependence $G(\xi)$ in \eqref{e31} may be either calculated with Eq.~\eqref{e7b}, or taken from simulations as the on-axis potential $\Phi(0,\xi)$ divided by calculated $R(0)$ and constants [see Eq.~\eqref{e6a}]. Both are in excellent agreement with the simulations.

\begin{figure}[tbh]
\includegraphics[width=194bp]{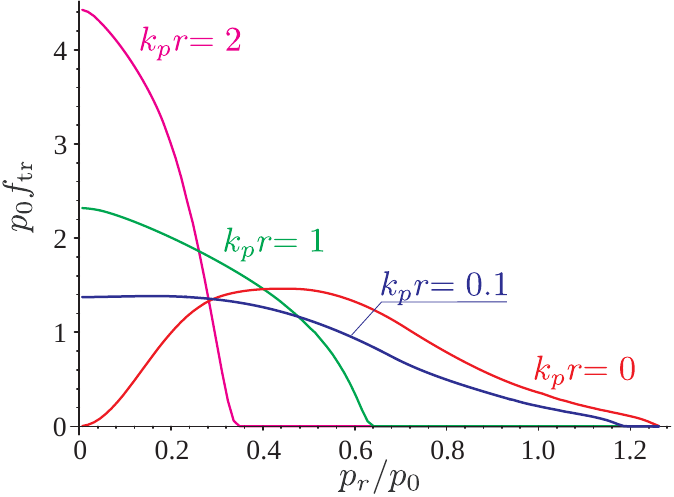}
\caption{ Transverse momentum distributions of beam particles $f_\text{tr}$ at different radii.}\label{fig11-distfunc}
\end{figure}

To calculate the transverse momentum distributions of beam particles at some radius $r$, we first need to relate the momentum $p_r$ that a particle has at this point and the initial oscillation amplitude of this particle $r_{a0}$. Combining \eqref{e29}, \eqref{e21}, and \eqref{e6a}, we obtain
\begin{equation}
\label{e33}
    r_{a0} (r, p_r) = \frac{r_a}{B(r_a)}, \quad
    r_a (r, p_r) = R^{-1} \left(
        R(r) + \frac{p_r^2}{p_0^2}
    \right)
\end{equation}
where $R^{-1}$ is the inverse function to the well shape. The number fraction of particles that have the radial momentum between $p_r$ and $p_r+dp_r$ is
\begin{equation}\label{e34a}
    D_0 (r_{a0}) \left.\frac{\partial r_{a0}}{\partial p_r}\right|_r dp_r.
\end{equation}
These particles spend the fraction
\begin{equation}\label{e35}
    \frac{dt}{\tau(r_a)} \propto \frac{dr}{p_r \tau(r_a)}
\end{equation}
of their time in the interval $dr$, with  $\tau$ and $dt$ determined by Eqs.~\eqref{e6} and \eqref{e7}. Thus, the momentum distribution is
\begin{equation}\label{e34}
    f_\text{tr} (r, p_r, \xi) = \frac{A(r) D_0 (r_{a0})}{p_r \tau (r_a)} \left.\frac{\partial r_{a0}}{\partial p_r}\right|_r,
\end{equation}
where $A(r)$ is a normalization constant. As follows from the derivation, for any radius $r$ the momentum scale of the distribution is determined by $p_0$, while the distribution shape is different at different $r$ (Fig.\,\ref{fig11-distfunc}). Note that the distribution is not Gaussian, unlike sometimes assumed by default.\cite{PFB2-1376,PRE49-4407} In particular, the maximum transverse momentum of beam particles at a given $r$ is
\begin{equation}\label{e37}
    p_{r,\text{max}} = \sqrt{2 \gamma_b m |\Phi(r, \xi)|}.
\end{equation}

\section{Summary}\label{s6}

An axisymmetric low-emittance charged particle beam in a dense plasma quickly reaches equilibrium with the wakefield it excites. Decelerated parts of the beam evolves towards some universal equilibrium state that is derived and described in this paper. For accelerated parts of the beam, no universal equilibrium state is found yet.

The universal equilibrium state is rather unusual. The beam density is strongly peaked near the axis and has a $1/r$ singularity. The radial electric field is approximately constant up to a small radius determined by the initial beam emittance. The beam radius is constant along the beam and equals approximately one half of the initial beam radius. The beam emittance varies along the beam in proportion to the square root of the potential well depth that confines beam particles radially. The frequency of transverse particle oscillations in this well strongly depends on the oscillation amplitude and, for most particles, is several times higher than the commonly used estimate. The transverse momentum distribution of beam particles depends on the observation radius and is not Gaussian.

\acknowledgements

This work is supported by The Russian Science Foundation, grant No.~14-12-00043. The computer simulations are made at Siberian Supercomputer Center SB RAS.

\end{document}